\documentclass[preprint]{revtex4}

\usepackage[utf8]{inputenc}
\usepackage[T1]{fontenc}
\usepackage[english,activeacute]{babel}
\usepackage{amsmath, amsthm, amsfonts, amssymb}
\usepackage{graphicx}

\newcommand{\nd}{\noindent}

 \theoremstyle{mdpi}
 \newcounter{thm}
 \newcounter{ex}
 \newcounter{re}
 \theoremstyle{mdpidefinition}

 \newtheorem{Definition}[thm]{Definition}

\def\Tr{\operatorname{Tr}}
\def\Salicru{H_{(h,\phi)}}
\def\SalicruQ{\mathbf{H}_{(h,\phi)}}
\def\SalicruG{\mathrm{H}_{(h,\phi)}}

\begin{document}

\title{Quantum information as a non-Kolmogorovian generalization of Shannon's theory}

\author{F. Holik, G.M. Bosyk and G. Bellomo}

\address{Instituto de F\'isica La Plata (IFLP), CONICET, and Departamento de F\'isica, Facultad de Ciencias Exactas, Universidad Nacional de La Plata, C.C. 67, 1900 La Plata, Argentina\\
}

\begin{abstract}
In this article we discuss the formal structure of a generalized information theory based on the extension of the probability calculus of Kolmogorov to a (possibly) non-commutative setting. By studying this framework, we argue that quantum information can be considered as a particular case of a huge family of non-commutative extensions of its classical counterpart. In any conceivable information theory, the possibility of dealing with different kinds of information measures plays a key role. Here, we generalize a notion of state spectrum, allowing us to introduce a majorization relation and a new family of generalized entropic measures.
\end{abstract}

\keywords{Quantum Probability; Generalized Majorization; Information Theory; Quantum Information}

\maketitle


\section{Introduction}

Quantum Information Theory is not only interesting because of its promising technological applications, but also by its impact at the very heart of physics, giving place to a new way of studying quantum mechanics~\cite{Bub-2005} and other possible theories as well. In particular, it has given rise to a quest for the foundational principles that singularize quantum mechanics among a vast family of possible statistical theories~\cite{PopescuNature-2014,Amarral-Cabello-EP,Cabello-EP-2014,CBH}. The study and characterization of quantum correlations plays a central role in this quest~\cite{Bruner-2014-ReviewBell}, being entanglement~\cite{Schro1,Schro2,werner89} and discord~\cite{zurek01} the most important ones. As is well known, probabilities and correlations are essential concepts in both classical and quantum information theories. But it turns out that the probabilities involved are fundamentally different on each of these theories. In this work, we will argue that, due to quantum contextuality and the non-Kolmogorovian nature of the underlying probabilities, quantum information theory can be correctly characterized as a non-commutative version of its classical counterpart.

For a statistical theory, it is very illuminating to look at the geometrical aspects of the set of possible states. This has been done extensively for the quantum case~\cite{Bengtsson2006,LudwigBookI,LudwigBookII,VaradarajanBookI,VaradarajanBookII,Mielnik-68,Mielnik-69,Mielnik-74,kibble1979}. But the geometry of a quantum set of states differs radically from that of a classical one. While the set of states of classical and quantum systems share the characteristic of being convex sets, alike quantum ones, classical models are simplexes. This difference expresses itself also at the level of the axiomatization. While Kolmogorov's axioms suffice for describing classical probabilistic models, the Boolean structure of a sigma algebra must be generalized to an orthomodular lattice of projection operators for the quantal case~\cite{BvN,dallachiaragiuntinilibro,mikloredeilibro,belcas81,svozillibro,HandbookofQL,jauch,piron,kalm83,kalm86,dvupulmlibro,VaradarajanBookI,VaradarajanBookII}.

One may wonder if there are probabilistic models more general than quantum and classical ones. This is indeed the case, and we must not go too far from standard quantum mechanics in order to find them. For example, in algebraic relativistic quantum field theory, states may be defined as measures over Type III factors~\cite{Yngvason2005-TypeIIIFactors}, a special kind of von Neumann algebras~\cite{RingsOfOperatorsI,RingsOfOperatorsII,RingsOfOperatorsIII,RingsOfOperatorsIV,vN-OriginalPaperOnVNAlgebras-1930,HalvorsonARQFT}, which differs from the Type I factors appearing in standard quantum mechanics~\cite{Redei-Summers2006,Redei-2012,RedeiHandbook,mikloredeilibro}. Type II factors can also be found in algebraic statistical mechanics (quantum mechanics with infinite degrees of freedom)~\cite{Redei-Summers2006,OlaBratteli}. Thus, it is clear that states defining measures which go beyond standard quantum mechanics exist, and they appear in examples of interest for physics. These new measures, which go beyond the distributive (or equivalently, commutative) case of the Boolean sigma algebra, are sometimes called non-Kolmogorovian or non-commutative probabilities. The fact that these non-commutative probabilities are involved \textit{is responsible for the emergence of the peculiar features of quantum information theory}.

But then, one could also imagine a setting where more general probabilistic theories can be conceived in order to study its general features and compare them with already known ones. This approach has been developed by many authors, and it is fair to say that it is based on the study of convex sets of states which define measures on certain algebras of observables. These are usually called events or, more generally, effects (see for example~\cite{Beltrametti.Varadarajan-2000,wilce,Barnum-Wilce-2009,Barnum-Wilce-2010,Barnum-PRL,Barnum-Toner,Entropy-generalized-II,Barret-2007,Perinotti-2011,GenealizedTeleportationBarnum,
Holik-Plastino-Massri}). The origins of this approach could be traced to the works of G. Ludwig~\cite{LudwigBookI,LudwigBookII} and G. Mackey~\cite{mackey-book}, but also to von Neumann. See also~\cite{Redei-Summers2006,Davies-Lewies,Srinivas,Gudder-StatisticalMethods,mackey-book} for other axiomatizations of non-Kolmogorovian probabilities and their relationships with lattice theory. Non-linear generalizations of quantum mechanics where studied using a similar approach in~\cite{Mielnik-68,Mielnik-69,Mielnik-74}.

It is possible to study many important notions of information theory such as entanglement, discord, and many information protocols in generalized probabilistic models (see for example~\cite{Entropy-generalized-II,Barret-2007,Barnum-Toner,Perinotti-2011,Holik-Plastino-Massri,GenealizedTeleportationBarnum,Barnum-PRL,Barnum-Wilce-2009,Barnum-Wilce-2010,ReviewNOVA}). We will argue in favour of the existence of a generalized information theory, continuing the lines of previous works~\cite{Entropy-generalized-II,Barret-2007,Holik-NaturalInformationMeasures} (see also~\cite{HolevoBook-2011} and~\cite{HolevoStatisticalStructure}, where non-commutative versions of many statistical techniques are studied). By focusing on the study of the formal aspects of the probabilities involved in different models, we show that the non-Kolmogorovian character of the probabilities underlying the quantum formalism is responsible for the emergence of quantum information theory\footnote{See also~\cite{Ingarden197643} for a discussion in which the role played by the formal structure of quantum probabilities in the begginigs of quantum information theory is stressed.}. This allows us to claim: \textit{Kolmogorovian probabilities imply Shannon's information theory; the non-commutative probability calculus of quantum theory, implies quantum information theory.} Quantum and classical information theories appear as particular instances of a formalism based on generalized probabilistic measures.

Any information theory depends strongly on our capability of dealing with different information measures. This is the case in classical information theory, where Shannon's~\cite{Shannon}, Tsallis~\cite{Tsallis1988} and R\'{e}nyi~\cite{Renyi1961} entropies (among other measures) are used for different purposes. A similar diversity of measures should be available in the generalized probabilistic setting. Previous works have focused in some entropic measures in the setting of generalized probabilities~\cite{Entropy-generalized-II,Barnum2010,HeinEntropyInQL1979}. In this paper we extend a new family of entropies based on the $(h,\phi)$-entropies to the general probabilistic setting~\cite{Salicru1993,Holik-Bosyk-Entropies}. These measures include the previous ones studied in the literature as particular cases. Another important notions introduced in this article are a definition of \textit{generalized spectrum} for states in general models, and a relationship of \textit{generalized majorization} between states. These are shown to be useful for defining functions of states and studying the properties of the entropic measures.

The paper is organized as follows. In Section~\ref{s:Kolmogorov} we review classical probabilities in the Kolmogorov approach. Next, we turn to important aspects of the quantum formalism and the formal structure of probability measures in quantum mechanics in Section~\ref{s:QuantumProbabilities}. In Section~\ref{s:COMpreliminaries} we discuss the formal aspects of a generalized information theory in the operational approach. In Section~\ref{s:GeneralizedEntropies}, we introduce our new family of information measures and the notion of \textit{generalized spectrum}, which allow us to introduce the concept of \textit{generalized majorization}. Finally, in Section~\ref{s:Conclusions} we draw our conclusions.

\section{Classical Probabilities}\label{s:Kolmogorov}

One of the most used axiomatizations of classical probability theory is the one of A.N. Kolmogorov~\cite{KolmogorovProbability}. If the possible outcomes of an experiment are represented by a set $\Omega$, subsets of it can be considered as representing events. It is usual to restrict events to a $\sigma$-algebra $\Sigma$ of subsets of $\Omega$. Thus, Kolmogorov defines probability measures as functions $\mu$ such that
    \begin{subequations} \label{e:kolmogorovian}
    \begin{equation}
    \mu:\Sigma\rightarrow[0,1] \,, \label{e:kolmogorovian1}
    \end{equation}
satisfying
    \begin{equation}
    \mu(\Omega)=1 \,, \\
    \end{equation}
and, for any pairwise disjoint denumerable family $\{A_{i}\}_{i\in I}$,
    \begin{equation}
    \mu(\bigcup_{i\in I}A_{i})=\sum_{i\in I}\mu(A_{i}) \,. \label{e:kolmogorovian3}
    \end{equation}
    \end{subequations}
In this way, Kolmogorov's approach puts probability theory in a direct connection with measure theory. From this axiomatic it is straightforward to see that $\mu(\emptyset)=0$ and $\mu(A^c)=1-\mu(A)$, where $(\cdot)^c$ means set-theoretical complement.

There exist many approaches to classical probabilities (see, e.g.~\cite{Rocchi-Probabilities} for a complete review). This subject is too vast to cover it here and goes beyond the scope of this work. We only mention the Bayesian school because of its importance and many physical applications ~\cite{CoxLibro,CoxPaper,Jaynes2} (see also~\cite{deFinetti}).


\section{Quantum Probabilities}\label{s:QuantumProbabilities}

In this Section we will discuss the special features of the probabilities involved in quantum theory. The most salient feature is that alike the classical case, the algebra of events of a quantum system is non-Boolean. This is related with the complementarity principle, for which incompatible experiments are needed to fully describe quantum phenomena.

\subsection{Elementary tests in quantum mechanics}

Propositions such as ``the value of the energy lies in the interval $(a,b)$'' or ``the particle is detected between the interval $(a,b)$'', are examples of how results of experiments can be expressed in quantum mechanics. Elementary propositions of that form are usually called \textit{events}, and they are represented by projection operators as follows. A projective valued measure (PVM) is a map $M$ such that
    \begin{subequations}
    \begin{equation}
    M: B(\mathbb{R})\rightarrow \mathcal{P}(\mathcal{H}) \,,
    \end{equation}
where $B(\mathbb{R})$ is any Borel set on $\mathbb{R}$ and $\mathcal{P}(\mathcal{H})$ is the space of projections on a Hilbert space $\mathcal{H}$, satisfying
    \begin{equation}
    M(\emptyset) = \mathbf{0} \ \mbox{and} \ M(\mathbb{R}) = \mathbf{1},
    \end{equation}
    where $\mathbf{0}$ is the null space and $\mathbf{1}$ the identity operator, and
    \begin{equation}
    M(\bigcup_{i\in I}B_{i})=\sum_{i\in I}M(B_{i}) \,,
    \end{equation}
    \end{subequations}
for any disjoint denumerable family $\{B_{i}\}_{i\in I}$. As in the classical case, from this axiomatic results that ${M(B^{c})=\mathbf{1}-M(B)} = M(B)^\bot$ (where $(\cdot)^\bot$ stands for orthogonal complement).

\vskip 3mm
\nd The spectral theorem allows to assign a PVM to any selfadjoint operator representing a physical observable $O$~\cite{dallachiaragiuntinilibro,mikloredeilibro} . We denote by $M_O$ its corresponding PVM. Thus, for any Borel set $(a,b)\in\mathbb{R}$ representing an interval of possible values of $O$, $M_O((a,b))=\textbf{P}_{(a,b)}$ is a projection operator that represents the elementary event "the value of $O$ lies in the interval $(a,b)$".

The state of a quantum mechanical system is represented by a density operator $\rho$, which is semi-definite positive and of trace one~\cite{Holik-Zuberman-2013}. Given $\rho$, the probability that the event represented by $\textbf{P}_{(a,b)}$ occurs is given by the Born's rule
    \begin{equation}\label{e:bornrule2}
    p(\textbf{P}_{(a,b)};\rho)=\Tr(\rho\textbf{P}_{(a,b)}) \,.
    \end{equation}

\noindent A generalization of the above mechanism for computing probabilities is given by the notion of \textit{quantal effects} and \emph{positive operator valued measures} (POVM))~\cite{Busch-Lahti-2009,Thesis-Heinonen-2005,Ma-Effects,EffectAlgebras-Foulis-2001,Cattaneo-Gudder-1999,bush,foulis}. In quantum mechanics, a POVM is represented by a mapping
    \begin{subequations}
    \begin{equation}
    E:B(\mathbb{R})\rightarrow\mathcal{B}(\mathcal{H}) \,,
    \end{equation}
where $\mathcal{B}(\mathcal{H})$ stands for bounded operator, such that
    \begin{gather}
    E(\mathbb{R})=\mathbf{1} \,, \\
    E(B)\geq 0, \,\,\mbox{for all}\,\, B\in B(\mathbb{R}) \,,   \\
    E(\bigcup_{i \in I}B_{i})=\sum_{i\in I}E(B_{i}),\,\, \mbox{for any disjoint family}\,\, \{B_{i}\}_{i \in I} \,.
    \end{gather}
    \end{subequations}

\noindent Then, the probability of effect $\textbf{E}$ given that the system is prepared in state $\rho$ is given by
\begin{equation}\label{e:bornrule3}
p(\textbf{E};\rho) = \Tr (\rho \mathbf{E}) \,.
\end{equation}

\subsection{Von Neumann's axioms}\label{s:vnAxioms}

Is there an analogous of Kolmogorov's axioms in quantum theory? As we have seen, events of a classical probabilistic theory can be represented as subsets of a given outcome set, yielding a Boolean $\sigma$-algebra. Consequently, classical states can be considered as measures over Boolean algebras. But as we have seen, the complementarity principle forces the non-commutativity of certain observables. This makes the algebra of projection operators (i.e., the algebra of possible events) non-distributive, and thus, non-Boolean. In this way, quantum states can be characterized as measures over non-Boolean algebras as follows:

\begin{subequations} \label{e:nonkolmogorov}
\begin{equation}
s:\mathcal{P}(\mathcal{H})\rightarrow [0,1] \,, \label{e:Qprobability1}
\end{equation}
such that
\begin{equation} \label{e:Qprobability2}
s(\mathbf{1})= 1, \\
\end{equation}
and, for a denumerable and pairwise orthogonal family of projections $\{\mathbf{P}_i\}_{i \in I}$,
\begin{equation} \label{e:Qprobability3}
s(\sum_{i \in I}\mathbf{P}_{i}) = \sum_{i \in I}s(\mathbf{P}_{i}) \,.
\end{equation}
\end{subequations}

\noindent We will refer to the above axioms as \textit{Kolmogorov's axioms}. Gleason's theorem~\cite{Gleason} asserts that the family of measures obeying von Neumann's axioms is in bijective correspondence with the set of positive trace class operators of trace one, which is nothing but the set of all possible quantum states. Thus, von Neumann's axioms relate quantum states with the non-Boolean (or non-commutative) measure theory defined by Eqs.~\eqref{e:Qprobability1}--\eqref{e:Qprobability3}. As remarked in the introduction, this fact lies behind the distinctive features of quantum information theory. Another important remark is that both the collection of all possible measures obeying von Neumann's axioms and the ones obeying Kolmogorov's form convex sets. This geometrical feature can be endowed of a natural physical interpretation: given two probability distributions, one can always form a mixture of them (and this will be represented mathematically by the corresponding convex combination in the state space).

\subsection{Quantum Correlations}

The non-abelian character of the quantum algebraic setting gives rise to a variety of new possibilities regarding correlations. So far, the most important of these novel quantum features has been the so called \textit{entanglement}. First recognized by Schrödinger and Einstein, Podolsky and Rosen in 1935, entanglement had remained in the centre of debate, inspiring discussions around the completeness of the formalism, the reality and locality of the theory, or, more recently, about its status as resource for quantum information processing tasks (see, e.g.,~\cite{horodecki07} for a complete review).

In the bipartite scenario, a quantum state is said non-entangled if and only if it can be approximated by convex linear combinations of product states. As Werner put it in his 1989 seminal paper, given a joint $AB$-bipartite state $\rho$, the state is \textit{separable} if there exist a probability distribution $\{p_k\}$ and marginal states $\{\rho^A_k\}$, $\{\rho^B_k\}$, such that~\cite{raggio88,werner89}

\begin{equation}
\rho = \sum_k p_k \rho^A_k \otimes \rho^B_k \,.
\end{equation}

\noindent Then, $\rho$ is entangled if it is not separable. This definition can be rephrased in more general algebraic terms. Let $\mathcal{N}_A$ and $\mathcal{N}_B$ be von Neumann algebras acting on a common Hilbert space, associated to the $A$ and $B$ subsystems. A state ${\omega_\rho:\mathcal{N}\rightarrow\mathbb{C}}$ is an expectation value functional, where ${\omega_\rho(n)=\text{Tr}(n\rho)}$ for any observable ${n\in\mathcal{N}}$. Then, ${\omega_\rho}$ on ${\mathcal{N}_A\vee\mathcal{N}_B}$ (the smallest von Neumann algebra generated by $\mathcal{N}_A$ and $\mathcal{N}_B$) is a \textit{product state} with respect to $\mathcal{N}_A$ and $\mathcal{N}_B$ iff ${\omega_\rho(ab)=\omega_\rho(a)\omega_\rho(b)}$ for any ${a\in\mathcal{N}_A}$ and any ${b\in\mathcal{N}_B}$. If $a$ and $b$ are projectors, ${\omega_\rho}$'s being a product state implies that the probability of measuring $ab$ factorizes ---the usual criterion for uncorrelation. Moreover, the state ${\omega_\rho}$ on ${\mathcal{N}_A\vee\mathcal{N}_B}$ is separable with respect to $\mathcal{N}_A$ and $\mathcal{N}_B$ iff it can be approximated by convex linear combinations of product states. Else, it is entangled.

As claimed before, the non-abelian nature of $\mathcal{N}_A$ ($\mathcal{N}_B$) is essential here. No entanglement is possible if the algebras are generated only by commutative observables~\cite{raggio88,earman14}. In other words: \textit{probabilities must be non-Kolmogorovian as a condition of possibility for true entanglement}. This fact has important consequences for quantum information processing, because entanglement plays a key role in the most useful protocols.

The non-commutativity is also responsible for the perturbation of the joint state when measuring over one of its parts. This fact can be quantified
by the difference between the pre and post-measurement mutual informations after a local (non-selective) measurement, a quantity known as \textit{discord}~\cite{zurek01,henderson01}. A non-discordant or \textit{classically-correlated} state $\rho$ is one that can be written as~\cite{datta10}
    \begin{equation}
    \rho = \sum_{ij} p_{ij} \Pi^A_i \otimes \Pi^B_j \,,
    \end{equation}
where $\{\Pi^A_i\}$ ($\{\Pi^B_j\}$) is a basis of orthogonal projectors on the Hilbert space of $A$ ($B$), and $\{p_{ij}\}$ is the corresponding probability distribution. Also, one can define states that are classically-correlated with respect to one of the parts only. For example, ${\rho=\sum_{ij} p_{ij} \Pi^A_i \otimes \rho^B_j}$ would be a classical-quantum state. Regarding its accessible information, a classical-quantum state can be locally measured in $A$ to obtain maximal information about the joint state without perturbing the same state. In the last decade, quantum discord was also identified with the quantum advantage for some informational tasks (see~\cite{modi12} for a complete review). Notice that in order to have non-null discord, non-orthogonal (i.e., incompatible) projections must be involved: this is another way in which the non-Boolean character of the event algebra is expressed.

As we explain below, the notions of entanglement and discord are susceptible to be extended upon general probabilistic theories.

Finally, it is worth to note that there are many other ways to assess the quantum peculiarities. For example, \textit{steering} ---first proposed by Schrödinger~\cite{Schro1}, and which has recently attracted a lot of attention~\cite{Witt12,Bran12,Smith12,Reid13,Jevt14}--- concerns the perturbation of a distant part trough the manipulation of local degrees of freedom, and is closely related to the notion of non-locality.

\section{Generalized Setting}\label{s:COMpreliminaries}

The lattice of projection operators of a separable Hilbert space and that of $\sigma$-algebras, are special instances of orthomodular lattices \cite{kalm83}. Orthomodular lattices are a suitable framework for describing contextual theories: given an orthomodular lattice $\mathcal{L}$, each possible context will be represented by a maximal Boolean subalgebra. If the maximal Boolean subalgebra coincides with the original lattice, then, the theory will be non-contextual. In order to describe theories more general than quantum mechanics, one could generalize the above axioms for probability theory to arbitrary orthomodular lattices as follows. Given $\mathcal{L}$, define a measure $\nu$ satifying

\begin{subequations}\label{e:GeneralizedProbability}
\begin{equation} \label{e:GenProb1}
    \nu:\ \mathcal{L} \rightarrow [0,1],
\end{equation}
such that
\begin{equation}
    \nu(\mathbf{1})=1, \\
\end{equation}
and, for a denumerable and pairwise orthogonal family of events $\{E_{i}\}_{i\in I}$,
\begin{equation} \label{e:GenProb3}
    \nu(\sum_{i\in I}E_{i}) = \sum_{i\in I}\nu(E_{i}) \,.
\end{equation}
\end{subequations}

\noindent See e.g.~\cite{belcas81} for conditions under which these measures exist. It is important to remark that Eqs.~\eqref{e:kolmogorovian1}--\eqref{e:kolmogorovian3} and~\eqref{e:Qprobability1}--\eqref{e:Qprobability3} are just particular examples of the above axioms. But these are much more general: in algebraic relativistic quantum field theory and in algebraic statistical mechanics more general orthomodular lattices appear~\cite{Redei-Summers2006,Yngvason2005-TypeIIIFactors,HalvorsonARQFT}. Many of the informational notions that can be described in quantum mechanics can be generalized to this formal setting (see for example \cite{HeinEntropyInQL1979}, \cite{HolikMaxEnt} and \cite{Holik-SymmetryMaxEnt}, where the Maximum Entropy principle is analyzed). It is also important to mention that other types of non-Kolmogorovian probabilistic theories can be conceived (we will not deal with them here, but see for example \cite{AcacioNegative} and \cite{CarnielliParaconsistent-2015}).

\vskip 0.3cm

In Section~\ref{s:vnAxioms}, we have mentioned that both quantum and classical state spaces are convex sets. This has to do with the fact that the collection of measures over an orthomodular lattice can be always endowed with a convex set structure (it is straightforward to show this for measures obeying axioms~\eqref{e:GenProb1}--\eqref{e:GenProb3}). The convex structure of the state space will play a key role in probabilistic theories.

\textit{Is it possible to describe a generalized probabilistic theories using convex sets as the starting point?} The answer is affirmative (see for example~\cite{Mielnik-68,Mielnik-69,Mielnik-74} and~\cite{Barnum-Wilce-2009,Barnum-Wilce-2010,ReviewNOVA,Barnum-PRL,Barnum-Toner}). Let us denote by $\mathcal{C}$ to the set of all possible states of an arbitrary model. It is reasonable to assume that $\mathcal{C}$ is convex, given the fact that we should be allowed to make mixtures of states. Given an observable quantity, denote by $X$ to the set of its possible measurement outcomes. Given an arbitrary state $\nu\in\mathcal{C}$ and any outcome $x\in X$, a number $\nu(x)\in[0,1]$ should be assigned, representing the probability of obtaining the outcome $x$ given that the system is prepared in state $\nu$. Using this, for outcome $x$ we can define an affine evaluation-functional $E_{x}\mathcal{C}\rightarrow [0,1]$ in a canonical way by $E_{x}(\nu)=\nu(x)$.

As $\mathcal{C}$ is convex, it can be naturally embedded in a vector space $V(\mathcal{C})$. Thus, any affine functional acting on $\mathcal{C}$ belongs to a dual space $V^{\ast}(\mathcal{C})$. It is very natural then to consider any affine functional $E:\mathcal{C}\rightarrow [0,1]$ as representing a possible measurement outcome or \textit{generalized effect}, and as above, to interpret $E(\nu)$ as the probability of finding the outcome represented by effect $E$ if the system is prepared in state $\nu$. It is very natural also to assume that there exists a normalization functional $u_{\mathcal{C}}$ such that $u_{\mathcal{C}}(\nu)=1$ for all $\nu\in\mathcal{C}$ (in the quantum case, this functional is represented by the trace functional). A (discrete) observable will be then represented by a set of effects $\{E_{i}\}$ such that $\sum_{i}E_{i}= u_{\mathcal{C}}$.

$\mathcal{C}$ will be said to be finite dimensional if and only if $V(\mathcal{C})$ is finite dimensional. In this paper, we will restrict for simplicity to this case and to compact sets of states. These conditions imply that $\mathcal{C}$ will be expressed as the convex hull of its extreme points. As in the quantum and classical cases, extreme points of the convex set of states will represent \emph{pure states}.

Define a finite dimensional  simplex as the convex hull of $d+1$ linearly independent points. A system is said to be classical if and only if it is a simplex. It is a well known fact that in a simplex a point may be expressed as a unique convex combination of its extreme points. This characteristic feature of classical theories no longer holds in quantum models. Indeed, in the case of quantum mechanics, there are infinite ways in which one can express a mixed as a convex combination of pure states (for a graphical representation, think about the maximally mixed state in the Bloch sphere).

\vskip 0.3cm

Interestingly enough, there is also a connection between the faces of the convex set of states of a given model and its lattice of properties (in the quantum-logical sense), providing an unexpected connection between geometry, lattice theory and statistical theories~\cite{Bengtsson2006,belcas81,HolikRingsOfOperators}. $F$ is a face if for all $x$ satisfying

\begin{equation}
x=\lambda x_1+(1-\lambda)x_2, \,\,\,\,\,0\leq\lambda\leq1 \,,
\end{equation}

\nd then $x\in F$ if and only if $x_1\in F$ and $x_2\in F$~\cite{Bengtsson2006}. Thus, faces of a convex set can be interpreted geometrically as subsets which are stable under mixing and purification. It is possible to show that the set of faces of any convex set can be endowed with a lattice structure in a canonical way. For a classical model (i.e., described by a simplex) it turns out that the lattice is Boolean. \textit{Thus, probabilities defined by clasical state spaces are Kolmogorovian}. On the other hand, in QM, the lattice of faces of the convex set of states (defined as the set of positive trace class hermitian operators of trace one), is isomorphic to the von Neumann lattice of closed subspaces $\mathcal{P}(\mathcal{H})$~\cite{Bengtsson2006,belcas81}. This is nothing but saying that quantum states obey von Neumann axioms. In this way, a clear connection can be made between the approach based on orthomodular lattices and the approach based on convex sets. A similar result holds for more general (but not all) state spaces, but we will not deal with this problem here (see \cite{belcas81} and \cite{Bengtsson2006} for more discussion on this subject).

\textit{It is very important to remark that general probabilistic models will fail to be Kolmogorovian in general. This has important consequences for the possible correlations that can be defined between different systems, and thus, for information theoretical purposes.}

\vskip 0.3cm

We mention finally an important remark about the different degrees of generality that can be attained using different frameworks. It is very reasonable to start with measures over orthomodular lattices, mainly because this framework includes an important family of physical examples (such as classical statistical theories, quantum mechanics, quantum statistics and relativistic quantum field theory), but also because it allows to represent complementarity in a very direct way. But more general models of interest can be constructed. For example, $\sigma$-orthomodular posets can be used as events algebras (by defining measures similarly as those defined by axioms~\ref{e:GeneralizedProbability}). All orthomodular lattices are $\sigma$-orthomodular posets, but the last ones are more general, because they can fail to be lattices~\cite{Gudder-StatisticalMethods}. Finally, the approach that uses convex sets as a starting point is more general that the one provided by orthomodular lattices (this is so because it is possible to find models for which no orthocomplementarity relation can be defined~\cite{belcas81}, and thus their lattice of faces fails to be orthomodular). Notwithstanding, in order to illustrate the most salient features of non-Kolmogorovian probabilistic models, it is sometimes sufficient to stay in the orthomodular lattices setting. This is what we will do mostly in this paper (but we will consider some more general examples in Section~\ref{s:GeneralizedEntropies}).

\subsection{Non-Kolmogorovian Information Theory and Contextuality}

Complementarity and contextuality \cite{AcacioContextuality-2015,AcacioContextuality-2015b,CabelloProposalFor-2010}, are salient features of quantum theory. The role of the complementarity principle in quantum information theory was discussed in~\cite{Renes-2013}, where it is shown that it is crucial for understanding the main features of quantum information protocols. One of the most important formal expressions of the complementarity principle is that of the non-commutativity of operators representing physical observables. And this is intimately connected with the non-Boolean structure of the lattice of projection operators. Furthermore, the success of the most important quantum computation algorithms is explained under the light of the projective geometry underlying the formalism of quantum theory in~\cite{BubQLandQI}.

To see how this contextual structure reappears in a more general setting, consider an orthomodular lattice $\mathcal{L}$. A maximal Boolean subalgebra is a subset $\mathcal{B}\subseteq\mathcal{L}$, such that: 1)  $\mathcal{B}$ is closed and is a Boolean algebra with respect to the operations inherited from $\mathcal{L}$ (i.e., it is a Boolean subalgebra) and 2) if $\mathcal{B}'$ is another Boolean subalgebra such that $\mathcal{B}\subseteq\mathcal{B}'$, then $\mathcal{B}=\mathcal{B}'$ (i.e., it is maximal). The important thing for us is that maximal Boolean subalgebras can be considered as representing particular experiments to perform on the system. To illustrate this point, think of a spin $\frac{1}{2}$ system. If we want to measure the spin component along axis $\hat{z}$, this will be represented by operator $\hat{\sigma}_{z}$. Then, this operator has associated the Boolean subalgebra $\{\mathbf{0},\arrowvert +\rangle\langle +\arrowvert,\arrowvert -\rangle\langle -\arrowvert^{\bot},\textbf{1}\}$, representing all possible events defined by the experiment which consists of measuring that quantity: spin up in direction $\hat{z}$ (``$\arrowvert +\rangle\langle +\arrowvert$''), spin down in direction $\hat{z}$ (``$\arrowvert -\rangle\langle -\arrowvert^{\bot}$''), the contradiction ``$\mathbf{0}$'' and the tautology ``$\textbf{1}$'' (which are the analogous of ``$\emptyset$'' and the whole outcome set ``$\Omega$'' in the classical case respectively).

Denote by $\textbf{B}$ to the set of all possible Boolean subalgebras of an orthomodular lattice $\mathcal{L}$. It is possible to show that $\mathcal{L}$ can be written as the sum of its maximal Boolean subalgebras~\cite{navara1991pasting},

\begin{equation}\label{e:BooleanAlgebrasPasting}
\mathcal{L}=\bigvee_{\mathcal{B}\in\textbf{B}}\mathcal{B} \,.
\end{equation}

\noindent What is the meaning of this technical result for generalized probabilistic theories? If $\mathcal{L}$ is Boolean, the result is trivial: the system can be described by using a single probability distribution over a single experimental setup. If it is not Boolean, it means that the event algebra of our theory may present mutually complementary contexts. In other words, we will need to perform incompatible experiments (each one represented by a maximal Boolean subalgebras) in order to fully describe phenomena. Notice that each generalized state $s$, when restricted to a maximal Boolean subalgebra $\mathcal{B}$, gives a Kolmogorovian probability measure $s\rvert_{\mathcal{B}}$. Taken together with Eq.~\eqref{e:BooleanAlgebrasPasting}, this implies that \textit{a generalized state on a contextual model can be considered as a collection of classical probabilities indexed by each empirical setup}. The generalized measure obeying Axioms~\eqref{e:GenProb1}--\eqref{e:GenProb3} provides a coherent pasting of this collection of Kolmogorovian measures. In the quantum case, this role is played by the density matrix representing the state of the system.

\vskip 0.3cm

These features can be taken as a starting point in the convex sets approach. For example, in~\cite{Barret-2007} (see also~\cite{Garner-2013} and~\cite{BruknerZeilinger-2009}), a state $s$ is considered as a list of probability distributions: ${s=(p(i,W))_{i=0,...,n-1;W=X_{0},....,X_{m-1}}}$. The possible $W$'s represent a set of fiducial measurements and the $i$'s label the outcomes of each measurement. Fiducial measurements represent sets of measurements out which the state can be determined. To fix ideas, let us look in detail at the qubit. In this case, each state can be specified as $s=(p(i,W))_{i=0,1;W=\hat{\sigma}_{x},\hat{\sigma}_{y},\hat{\sigma}_{z}}$. The observables represented by ${\hat{\sigma}_{x},\hat{\sigma}_{y},\hat{\sigma}_{z}}$ are sufficient to determine completely the state (i.e., they form a \textit{fiducial set}). Notice that from this perspective, a state is considered again as a collection of classical probability distributions.

\vspace{.3cm}

Non-Kolmogorovian probabilities are a condition of possibility for a departure from Shannon's classical information theory. This can be understood in a simple way following a generalization of the R. T. Cox approach to probability theory as follows:

\begin{itemize}
 \item R.T. Cox \cite{CoxPaper,CoxLibro} showed that if a rational agent is confronted with a Boolean algebra representing empirical events, then, any function measuring his degree of belief on the occurrence of any event must be equivalent to a classical probability calculus.
 In~\cite{HolikAnnals-2014}, it is shown that if a rational agent is confronted with a non-distributive algebra of physical events, then, the consistent probabilities must be those of Equations~\ref{e:GeneralizedProbability}.
 \item In a similar way, in the Cox approach it is shown that Shannon's entropy is the more natural information measure for an agent confronted with a Boolean algebra of events. But in~\cite{Holik-NaturalInformationMeasures}, it is shown that if the algebra is replaced by a non-Boolean one, then von Neumann and Measurement entropies must be used. In other words: if the event algebras are non-Boolean, then probabilities must be non-Kolmogorovian, and information measures depart from Shanon's entropy and more general classical ones (we will discuss the specific form of this departure in Section~\ref{s:GeneralizedEntropies}).
\end{itemize}

\noindent This is expressed clearly in the formal structure of classical and quantum information theories as follows. In Shannon's theory, a source emits different messages $x$ of an outcome set $X$ with probabilities $p_{x}$: this means that the probabilities involved are nothing but a Kolmogorovian measure over the Boolean algebra generated by the possible outcomes of the source. This implies that Shanon's entropy will play a key role in the formalism. For example, in the noiseless-channel coding theorem, the value of the Shannon's entropy of the source $H(X)$ measures the optimal compression for the source messages~\cite{Holik-WISI}. What changes in the quantum setting? Due to the fact that the final output of the source is now represented by a density matrix $\rho=\sum p_{x}\rho_{x}$ (i.e., by a non-Kolmogorovian measure), then, the von Neumann's entropy comes into stage. This is expressed for example, in Schumacher's quantum coding theorem, in which the optimal bound for coding is expressed in terms of this quantity~\cite{Schumacher-QuantumCoding-1995,Holik-WIQI}.

The role of the non-Kolmogorovian probability involved in the quantum state emitted by the source is also expressed in the existence of the Holevo bound: the mutual information between emitter and receiver will be bounded from above by a quantity depending on the von Neumann's entropy $S(\rho)$

\begin{equation}
I(X:Y)\leqslant S(\rho) -\sum p_{x} S(\rho_{x} )
\end{equation}

\noindent where $I(X:Y)$ represents the classical mutual information between random variables $X$ and $Y$. The above bound means that there is an intrinsic limit to the information accessible to the receiver. For example, it can be shown that if the original mixture is formed by non-orthogonal states, the Holevo bound implies that $I(X:Y)$ is strictly less than $H(X)$ (the Shannon's measure of the source), and then, it is impossible for the receiver to determine $X$ perfectly if he measures the observable $Y$~\cite{nielsen2010quantum}. This implies that if the states prepared by the emitter are non-orthogonal, it will not be possible for the receiver to determine the emitted state with certainty. This impossibility is directly related to the complementarity principle, and thus, to the non-Kolmogorovian character of the emitted quantum state.

\subsection{Communication And Correlations In the Generalized Setting}

\textit{Communication} is a central aspect of any possible kind of information theory. But communication involves more than one party: a message (or something) must be sent from one party to another. This is why the study of correlations is so important in order to account for the probabilistic aspects of a source. In order to show that informational notions can be studied in the general setting described above, a suitable description of multipartite states and correlations is needed. This has indeed been done quite extensively~\cite{Barnum-Wilce-2009,Barnum-Wilce-2010,Barnum-PRL,Barnum-Toner}, and many notions essential to quantum information processing (such as entanglement, no-cloning, no-boradcasting and teleportation), can be generalized and studied in arbitrary statistical models. A departure of classical information theory will be found in state spaces for which non-classical probabilities and correlations are involved, and we will review how this is directly related to the non-Kolmogorovian structure of the state space.

Let us consider a compound system, formed of parties $A$ and $B$, with state spaces $\mathcal{C}_{A}$ and $\mathcal{C}_{B}$ respectively. The joint system will also have a state space, let us denote it by $\mathcal{C}_{A}\otimes\mathcal{C}_{B}$ (the meaning of this notation is clarified below). In order to study its mathematical features, let us suppose that $\mathcal{C}_{A}\otimes\mathcal{C}_{B}$ can be included in the linear span of $V(\mathcal{C}_{A})\otimes V(\mathcal{C}_{B})$ (this assumption is discussed in \cite{Barnum-PRL}).  Consider the set which contains all bilinear functionals $\varphi:V(\mathcal{C}_{A})\times V(\mathcal{C}_{B})\longrightarrow\mathbb{R}$ satisfying $\varphi(E,E')\geq 0$ for all effects $E$ and $E'$ and $\varphi(u_{A},u_{B})=1$. It is very reasonable then, to call this set a maximal tensor product state space $\mathcal{C}_{A}\otimes_{max}\mathcal{C}_{B}$ for $A$ and $B$. $\mathcal{C}_{A}\otimes_{max}\mathcal{C}_{B}$ has the property of being the biggest set of states in $(V(\mathcal{C}_{A})\otimes V(\mathcal{C}_{B}))^{\ast}$ which assigns probabilities to all product measurements.

Analogously, a minimal tensor product state space $\mathcal{C}_{A}\otimes_{min}\mathcal{C}_{B}$ can be defined as the convex hull of all product states. This will be the analogous of the convex set of separable states in quantum mechanics (see \cite{Holik-Plastino-Massri} for more discussion on this). We will write a product state as $\nu_{A}\otimes\nu_{B}$ satisfying

\begin{equation}
\nu_{A}\otimes\nu_{B}(E,E')=\nu_{A}(E)\nu_{B}(E') \,,
\end{equation}

\noindent for all pairs $(E,E')\in V^{\ast}(\mathcal{C}_{A})\times V^{\ast}(\mathcal{C}_{B})$. Given these two extreme possibilities (maximal and minimal tnesor product state spaces), the set of states $\mathcal{C}_{A}\otimes\mathcal{C}_{B}$ of an actual model lies somewhere ``in between'':

\begin{equation}
    \mathcal{C}_{A}\otimes_{min}\mathcal{C}_{B}\subseteq\mathcal{C}_{A}\otimes\mathcal{C}_{B}\subseteq\mathcal{C}_{A}\otimes_{max}\mathcal{C}_{B} \,.
    \end{equation}

\noindent For classical compound systems (for which state spaces are simplices representing Kolmogorovian probabilities), the set of compound states equals to the minimal tensor product (and is again a classical state space). This means that if both subsystems are classical, we recover the equality: $\mathcal{C}_{A}\otimes_{min}\mathcal{C}_{B}=\mathcal{C}_{A}\otimes_{max}\mathcal{C}_{B}$. It can be shown that for quantum mechanics we have the strict inclusions $\mathcal{C}_{A}\otimes_{min}\mathcal{C}_{B}\subseteq\mathcal{C}_{A}\otimes\mathcal{C}_{B}\subseteq\mathcal{C}_{A}\otimes_{max}\mathcal{C}_{B}$.

With this formal setting, it is now very natural to introduce a general definition of separable state in an arbitrary convex operational model. This is done in an analogous way to that of~\cite{werner89} (see for example \cite{Barnum-Toner,Perinotti-2011}:

\begin{Definition}\label{d:generalseparable}
A state $\nu\in\mathcal{C}_{A}\otimes\mathcal{C}_{B}$ will be called \emph{separable} if there exist $p_{i}\in\mathbb{R}_{\geq 0}$, $\nu^{i}_{A}\in\mathcal{C}_{A}$ and $\nu^{i}_{B}\in\mathcal{C}_{B}$ such that
\begin{equation}
\nu=\sum_{i}p_{i}\nu^{i}_{A}\otimes\nu^{i}_{B},\quad \sum_i p_i =1 \,.
\end{equation}
\end{Definition}

\noindent Entangled states are thus defined as those which are not separable. It can be easily checked that entangled states exist if and only if $\mathcal{C}_{A}\otimes\mathcal{C}_{B}$ is strictly greater than $\mathcal{C}_{A}\otimes_{min}\mathcal{C}_{B}$. \textit{Thus, no entangled states exist for classical theories. In this way, non-classical correlations will not be allowed, and no departure of classical information theory will be found.}

It is worth noting that this generalization of entanglement, although natural, is by no way unique, neither the most general possibility. In~\cite{barnum03-ent,barnum04-ent,barnum05-ent} Barnum \textit{et al.} propose a subsystem-independent concept of entanglement, where the focus is in the relation between the convex set of states and a preferred (relevant or prescribed by any means) set of effects. Then, entanglement becomes as a relative notion of purity of the states with respect to the relevant effects (see~\cite{barnum03-ent,barnum05-ent,viola10-ent} for details). Being independent of a certain subsystem decomposition, this notion becomes substantially more general than the usual one, even in the quantum scenario (see, e.g.~\cite{somma04,benatti10,balachandran13,balachandran13b}).

\vspace{.3cm}

Regarding discord, Perinotti studied a possible introduction of the notion in general probabilistic theories~\cite{Perinotti-2011}. As the original definitions of quantum discord relies on the information content of the states, and because the information measures are not uniquely defined for general probabilistic theories, Perinotti prefers to give an operational definition of discord. He starts by defining the set of null-discord states and proves that they can be expressed as

\begin{equation}
\omega_{nd} = \sum_{i\in I} q_k (\psi_A^i\otimes\sigma_B^i) \,,
\end{equation}

\noindent where ${\{\psi_A^i\}_{i\in I}}$ is a set of jointly perfectly distinguishable pure states, ${\{\omega_B^i\}_{i\in I}}$ is a set of arbitrary states of $B$, and ${\{q_i\}_{i\in I}}$ is a probability distribution (see~\cite{Perinotti-2011} for details). Then, the discord of a state $\nu$ is defined as the minimal \textit{operational distance} to the set of null-discord states ${\Omega_{nd}}$:

\begin{equation}
\mathcal{D}(\nu) := \min_{\omega_{nd}\in\Omega_{nd}}||\nu-\omega_{nd}||_{op} \,.
\end{equation}

\noindent The operational distance is defined through the minimum error probability in discrimination of both states~\cite{Chiribella-2011}.

\vskip 0.5cm

The fact that correlations between different parties can be studied using information measures in the generalized setting, allows to pose the problem of communication in a suitable mathematical form. Given that the probabilistic models involved can be non-Kolmogorovian, the departure from Shannon's formalism is unavoidable in most cases.

\section{Generalized Entropies And Majorization} \label{s:GeneralizedEntropies}

In this Section we extend the definition of classical and quantum Salicr\'u entropies to the case of general probabilistic theories. In addition, we introduce definitions of spectra of states and majorization in those theories.

\subsection{Entropies and majorization in classical and quantum theories}

Inspired in~\cite{Csiszar1967}, Salicr\'u \textit{et al.} have introduced a very general expression for entropies~\cite{Salicru1993}, which we call as classical $(h,\phi)$-entropies, as follows

\begin{Definition} \label{def:ClassicalSalicru}
For an $N$-dimensional probability vector $p = \{p_i\}$ with $p_i \geq 0$ and ${\sum_{i=1}^N p_i=1}$, the classical $(h,\phi)$-entropies are defined as
  \begin{equation} \label{eq:SalicruEnt}
  \Salicru(p) = h\left( \sum_{i=1}^N \phi(p_i) \right) \,,
  \end{equation}
where \textit{entropic functionals} $h:R\mapsto R$ and $\phi:[0,1]\mapsto R$ are continuous with $\phi(0)=0$ and $h(\phi(1))=0$, and are such that either: (i) $h$ is increasing and $\phi$ is concave, or (ii) $h$ is decreasing and $\phi$ is convex.
\end{Definition}

\nd It is straightforward to see that this definition yields the most renowned entropies, namely Shannon~\cite{Shannon}, Tsallis~\cite{Tsallis1988} and R\'enyi ones~\cite{Renyi1961} as particular cases. Indeed, one key property that all these entropies share is related to the concept of majorization~\cite{MarshallBook}. Majorization gives a partial order between probability vectors and it is defined as follows: for given probability vectors $p$ and $q$ of length $N$ sorted in decreasing order, it is said that $p$ is majorized by $q$, denoted as $p \prec q$, when
    \begin{equation} \label{e:majorization}
    \sum_{i=1}^n p_i \leq \sum_{i=1}^n q_i \ \mbox{for all} \ n=1, \ldots , N-1 \ \mbox{and} \ \sum_{i=1}^N p_i = \sum_{i=1}^N q_i.
    \end{equation}

In~\cite{Holik-Bosyk-Entropies,Zozor2014}, it has been shown that classical $(h,\phi)$-entropies are Schur--concave, that is, preserve the majorization relation: if $p \prec q \Rightarrow H_{(h,\phi)}(p) \geq H_{(h,\phi)}(q)$. Many properties of Salicr\'u entropies can be proved using majorization, e.g the lower and upper bounds: $0 \leq H_{(h,\phi)}(p) \leq h\left(N \phi\left( \frac{1}{N} \right)\right)$.

On the other hand, it is quite natural to define a quantum $(h,\phi)$-entropies replacing probability vector by density operator and the sum by the trace in Def.~\ref{def:ClassicalSalicru}, as follows~\cite{Holik-Bosyk-Entropies}

\begin{Definition} \label{def:QuantumSalicru}
Let us consider a quantum system described by a density operator $\rho$ acting on an $N$-dimensional Hilbert space $\mathcal{H}$. The quantum $(h,\phi)$-entropies (under the same assumptions for $h$ and $\phi$ in Def.~\ref{def:ClassicalSalicru}) are defined as follows
  \begin{equation} \label{eq:QuantumSalicru}
  \SalicruQ(\rho) = h\left( \Tr \phi(\rho) \right) \,.
  \end{equation}
\end{Definition}

As in the classical counterpart, the quantum $(h,\phi)$-entropies include as particular cases von Neumann~\cite{vonNeumann1927}, and quantum version of R\'enyi and Tsallis entropies. It can be shown that if the probability vector $p$ is formed by the eigenvalues of $\rho$, then
    \begin{equation} \label{e:equivCQSalicru}
    \SalicruQ(\rho) = \Salicru(p) \,.
    \end{equation}
In other words, quantum $(h,\phi)$-entropies are nothing more than classical $(h,\phi)$-entropies of the probability vectors formed by eigenvalues of density operators.

Let us consider two density operators $\rho$ and $\sigma$ with $p$ and $q$ vectors formed by eigenvalues sorted in decreasing order, respectively. Now, $\rho$ is majorized by $\sigma$, denoted as $\rho\prec\sigma$, means that $p \prec q$ in the sense of Eq.~\eqref{e:majorization}. It can be shown that quantum $(h,\phi)$-entropies are also Schur-concave~\cite{Holik-Bosyk-Entropies}.

Let $p(E;\rho)$ be the probability vector whose components are given by the Born rule for a rank-one POVM $E$ and state $\rho$, that is $p_i(\mathbf{E}_i;\rho)= \Tr \rho \mathbf{E}_i$. An alternative definition of quantum $(h,\phi)$-entropies, which is equivalent to Def.~\ref{def:QuantumSalicru} but with more physical meaning related to the probability of measurement, is the following~\cite{Holik-Bosyk-Entropies}:

\begin{Definition} \label{def:QuantumSalicru2}
Under the same assumptions in Def.~\ref{def:QuantumSalicru}, the quantum $(h,\phi)$-entropies are also defined as
    \begin{equation} \label{eq:QuantumSalicru2}
    \SalicruQ(\rho) = \min_{E \in \mathbb{E}} \Salicru(p(E;\rho)) \,,
    \end{equation}
where $\mathbb{E}$ is the set of all rank-one POVMs.
\end{Definition}

Further properties of classical and quantum $(h,\phi)$-entropies are given in~\cite{Holik-Bosyk-Entropies} (and references there in).

\subsection{Entropies and majorization in general probabilistic theories}

Now, we aim to extend the definition of $(h,\phi)$-entropies to more general probabilistic theories. It is possible to do this at least in two different ways. First, one could start with an atomic orthomodular lattice $\mathcal{L}$ defining an algebra of events. A frame in $\mathcal{L}$ will be an orthogonal set $\{a_{i}\}_{i\in I}$ of atoms such that $\bigvee_{i \in I} a_{i}= \mathbf{1}$. Frames represent maximal experiments. For example, in quantum mechanics, any orthonormal basis (or rank-one PVMs) is a frame. Thus, for each frame $\mathcal{F}=\{a_{i}\}_{i\in I}$ and each state $\nu$, we have $p_{i} = \nu(a_{i})$. Then, $\{p_{i}\}$ defines a probability vector and this allows us to define the $(h,\phi)$-entropies relative to that frame
    \begin{equation} \label{e:FrameSalicruEntropies}
    H_{(h,\phi)}\rvert_{\mathcal{F}}(\nu)= h\left( \sum_{i \in I}\phi(\nu(a_{i})) \right) \,.
    \end{equation}
In order to give a definition independent of the frame, we have to take the minimum over all possible frames:

\begin{Definition} \label{def:GeneralizedSalicru1}
Let us consider a state $\nu \in \mathcal{C}$. The general $(h,\phi)$-entropies (under the same assumptions for $h$ and $\phi$ in Def.~\ref{def:ClassicalSalicru}) are defined as follows
    \begin{equation} \label{eq:GeneralizedSalicru1}
    \SalicruG (\nu) =\inf_{\mathcal{F} \in \mathbb{F}} H_{(h,\phi)}\rvert_{\mathcal{F}}(\nu) \,
    \end{equation}
\end{Definition}

\noindent where $\mathbb{F}$ is the set of all frames. This is the canonical way in which entropies can be defined in general probabilistic theories. We observe that this approach resembles Def.~\ref{def:QuantumSalicru2} for the quantum case. Measurement entropy given in~\cite{Barnum2010,Entropy-generalized-II} is a particular case of this approach. But it also includes other quantities, such as R\'{e}nyi and Tsallis in the case of general probabilistic theories. Notice that by taking the minimum over all possible frames, the contextual structure of the probability measures involved is made explicit.

There is another possible way in which $(h,\phi)$-entropies in quite general probabilistic theories can be defined: we will provide a generalization of Def.~\ref{def:QuantumSalicru}. For this task, we have to define the notions of generalized spectrum and majorization. We restrict to arbitrary (compact) convex sets of states in finite dimensions: for these spaces, each element can be written as a convex combination of its pure states (as is the case in quantum and classical mechanics). In other words, there exist pure states $\{\nu_{i}\}$ such that every state $\nu$ can be written as

\begin{equation} \label{e:Decomposition}
\nu= \sum_{i}p_{i}\nu_{i} \,.
\end{equation}

\noindent But this decomposition is not, in general, unique. For instance, the maximally mixed state in quantum mechanics has infinite decompositions even in terms of orthogonal pure states. Notwithstanding, the probability vectors defined by the coefficients of these decompositions are all the same. Notice that this uniqueness property, needs not to be true for arbitrary models as we will discuss below.

We introduce now our notion of \textit{generalized spectrum} inspired in the Schr\"odinger mixture theorem (see e.g.~\cite[Th.8.2]{Bengtsson2006}). Using this theorem, it can be shown that the probability vector formed by the coefficients of any convex pure decomposition of a quantum state is majorized by the one formed by its eigenvalues. In other words, the spectrum of a quantum state has the distinctive property of being the majorant of all possible probability vectors originated in convex decompositions in terms of pure states. We will abstract this property, and use it for defining a generalized spectrum for generalized states as follows. Given a probabilistic model described by a compact convex set, let $M_{\nu}$ be the set of probability vectors of all possible convex decompositions of a state $\nu$ in terms of pure states, that is

\begin{equation}
M_{\nu}:=\{p(\nu)=\{p_i\} \,\arrowvert\,\nu=\sum_{i}p_{i}\nu_{i}\,\, \mbox{for pure}\,\,\nu_{i}\} \,.
\end{equation}

\noindent Then, we propose the following

\begin{Definition} \label{def:GeneralizedSpectrum}
Given a state $\nu$, if the majorant of the set $M_{\nu}$ (partially ordered by majorization) exists, it is called the spectrum of $\nu$ and it is denoted by $\bar{p}(\nu)$ .
\end{Definition}

\noindent Accordingly, the corresponding generalized spectral decomposition is

\begin{equation}
\label{def:GeneralizedSpectralDecomposition}
\nu = \sum_{i} \bar{p}_{i} \bar{\nu}_{i}.
\end{equation}

Notice that our definition reduces to the usual one for classical theories (where the sets of states are simplexes) and also in quantum mechanics. In the former case, equivalence can be checked easily, because there is only one convex decomposition in terms of pure states. In the latter case, as noted above, equivalence is a consequence of the Schr\"odinger mixture theorem. Notice however, that for a general statistical theory described by a compact convex set, it could be that the supremum $\bar{p}(\nu)$ does not exist for all possible states.

\begin{figure}
\centering
\includegraphics[width=.85\textwidth]{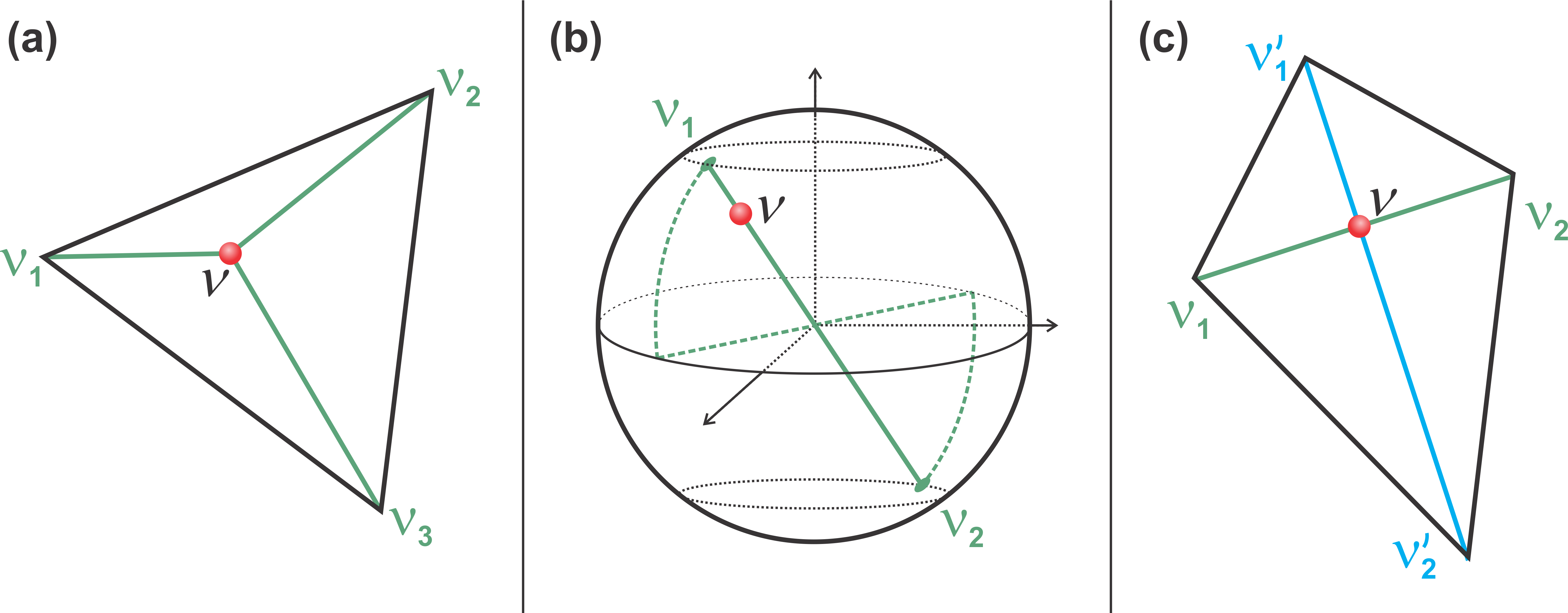}
\caption{The generalized spectral decomposition ---Eq.~\eqref{def:GeneralizedSpectralDecomposition}--- can be computed in a variety of probabilistic theories. \textbf{(a)} When the convex set is a simplex, the decomposition in terms of pure states is unique and so it determines the spectrum of $\nu$. In the triangle above, $\nu$ can be written in a unique way as a mixture of $\nu_1$, $\nu_2$ and $\nu_3$. \textbf{(b)} For the state $\nu$ of a qubit, the spectrum is given by the eigendecomposition of its density matrix in terms of the orthogonal pure states $\nu_1$ and $\nu_2$. The same happens for any other quantum mechanical model. \textbf{(c)} For a general probabilistic theory, there are, a priory, many decompositions of a state in terms of pure ones, and we have to look for the majorant one. For example, for the non-regular polygon with four vertices the state in the barycenter is ${\nu=\frac{1}{2}\nu_1+\frac{1}{2}\nu_2=x\nu_1'+(1-x)\nu_2'}$, with $x>\frac{1}{2}$. The second set of coefficients majorize the first one, so $\bar{p}(\nu)=\{x,1-x\}$ constitute the spectrum of $\nu$. Note, however, that in both decompositions the pure states are perfectly distinguishable.}
\label{fig:cases}
\end{figure}

We observe that an alternative definition of generalized spectrum has been recently introduced by Barnum \textit{et al.} in~\cite{Barnum-2015}. The authors define the spectrum of a state as the unique (up to permutations) convex decomposition into perfectly distinguishable pure states. Distinguishability has the following operational meaning: a set of states $\{\nu_i\}$ is perfectly distinguishable if there is a measurement $\{ \mathbf{E}_i \}$ such that $\mathbf{E}_i(\nu_j) =\delta_{ij}$. It is important to remark that their definition of spectrum cannot be used in arbitrary state spaces. This is due to the fact that for certain spaces, the decomposition of a state into perfectly distinguishable pure states can fail to be unique, and different decompositions can yield different probability vectors. Spaces for which decomposition into perfectly distinguishable states always exist, are said to satisfy the \textit{weak spectrality} axiom (WS-spaces). In spaces satisfying strong spectrality (S-spaces), the probability vectors of the convex pure decomposition into perfectly distinguishable states are unique (up to permutations). It can be shown that there are WS-spaces which are not S-spaces, and then, the definition of spectrality presented in~\cite{Barnum-2015} doesn't works in those cases.
The definition presented in~\cite{Barnum-2015} and ours yield the same result for classical and quantum state spaces. But they are expected to be non-equivalent in the general case. There could be spaces for which certain states admit different probability vectors for distinct decompositions into perfectly distinguishable pure states, but for which it is still possible to find a maximum according to our definition (see for example Fig.~\ref{fig:cases}). It is an interesting open question to determine under which conditions both definitions are equivalent, and specially, the range of validity of Def.~\ref{def:GeneralizedSpectrum}. This last task can be rephrased as follows: which are the spaces for which a generalized version of the Schr\"{o}dinger mixture theorem is valid? We will not deal with this problem here; we will only restrict to show how our definition can be used to define generalized majorization, functions over states and, in particular, entropic measures.

Def.~\ref{def:GeneralizedSpectrum} can be used to introduce naturally the concept of generalized majorization as follows.

\begin{Definition} \label{def:GeneralizedMajorization}
Given two states $\mu$ and $\nu$, one has that $\mu$ is majorized by $\nu$, denoted by $\mu\prec\nu$,  if and only if
\begin{equation}
\bar{p}(\mu) \prec \bar{p}(\nu) \,,
\end{equation}
where $\bar{p}(\mu)$ and $\bar{p}(\nu)$ are the corresponding generalized spectra from Def.~\ref{def:GeneralizedSpectrum}.
\end{Definition}

Moreover, our definition of generalized spectrum can be also used to evaluate a function $\phi$ in a generalized state as follows. For any possible mixture $\{p_i,\nu_i\}$ of $\nu$, we define the application of a functional $\phi$ to the state given the mixture as

\begin{equation}
\phi(\nu)\arrowvert_{\{p_{i}, \nu_i \}}:=\sum_{i}\phi(p_{i})\nu_{i} \,.
\end{equation}

\noindent In particular, we are interested in the mixture ${\{\bar{p}_{i}, \bar{\nu}_i \}}$, which leads to the definition

\begin{equation}
\phi(\nu):= \phi(\nu)\arrowvert_{\{\bar{p}_{i}, \bar{\nu}_i \}}\,.
\end{equation}

\noindent We have seen in Section~\ref{s:COMpreliminaries} that the partial trace of the quantum formalism can be extended to the general setting by using the normalization functional $u_{\mathcal{C}}$. This allow us to define alternative generalized $(h,\phi)$-entropies.

\begin{Definition} \label{def:GeneralizedSalicru2}
Under the same assumptions that in Def.~\ref{def:GeneralizedSalicru1}, we define the $(h,\phi)$-entropies
\begin{equation} \label{eq:fEntGeneralized}
\mathrm{\widetilde{H}}_{(h,\phi)}(\nu) = h\left( u_{\mathcal{C}} \,( \phi(\nu) )\right).
\end{equation}
\end{Definition}

\noindent In other words, these generalized entropies are are equal to the classical ones evaluated on the probability vector $\bar{p}(\nu)$, that is

$$ \mathrm{\widetilde{H}}_{(h,\phi)}(\nu) = \Salicru(\bar{p}(\nu))$$

\noindent In principle, it can be shown that all the properties of classical (and quantum) $(h,\phi)$-entropies that are based on majorization and Schur-concavity holds in this general case (further properties are under investigation).

\section{Conclusions}\label{s:Conclusions}

In this paper we have discussed the formal aspects that show that quantum information theory arises as a non-Kolmogorovian version of Shannon's information theory. In other words, when the probabilities involved are measures over projection lattices of Hilbert spaces, we obtain quantum information theory. On the other hand, when the algebra of events is a Boolean one, we recover Shannon's formalism. This structure is reencountered in the generalized setting, where many informational notions, such as correlations between different parties and information protocols can be described. In this way, quantum and classical information theories appear as particular cases of a generalized non-Kolmogorovian probabilistic calculus. In particular, we have shown that the Salicrú entropies can be defined in the non-Kolmogorovian setting, extending the catalogue of extant entropic measures available in the literature. In doing so, we have also introduced a definition of spectrum for generalized measures which relies in an essential property derived from the Schr\"{o}dinger mixture theorem, and which allows to introduce a new notion of generalized majorization and functions of states (such as the generalized entropies introduced in Def.~\ref{def:GeneralizedSalicru2}).

\acknowledgments

The Authors acknowledge CONICET and UNLP (Argentina). We are also grateful to the participants of the workshop What is quantum information? (Buenos Aires, May of 2015), Jeffrey Bub, Adán Cabello, Dennis Dieks, Armond Duwell, Christopher Fuchs, Robert Spekkens and Christopher Timpson, for the stimulating and lively discussions about the concept of information. This paper was partially supported by a Large Grant of the Foundational Questions Institute (FQXi), and by a grant of the National Council of Scientific and Technological Research (CONICET) of Argentina.

\bibliography{refs}
\bibliographystyle{apsrev}

\end{document}